\documentclass[fleqn,10pt]{wlscirep}  
\title{Elemental Phosphorus: structural and superconducting phase diagram under pressure} 

\newcommand{\tc}{T$_{\textmd C }$}   

\newcommand{\omlog}{$\omega_{\textmd log}$}

\newcommand{\sh}{SH$_3$}
\newcommand{\ph}{PH$_3$}

\newcommand{\simpcub}{$sc$} 
\newcommand{\hex}{$sh$}     
\newcommand{\rhomb}{$A7$}   
\newcommand{\Iftd}{$I-43d$} 
\newcommand{\bcc}{$bcc$}    

\author[1,*]{Jos\'e~A. Flores-Livas}
\author[2]{Antonio Sanna}
\author[3]{Alexander P. Drozdov}
\author[4]{Lilia Boeri}
\author[5]{Gianni Profeta}
\author[3]{Mikhail Eremets}
\author[1]{Stefan Goedecker}

\affil[1]{Department of Physics, Universit\"at Basel, Klingelbergstr. 82, 4056 Basel, Switzerland}
\affil[2]{Max-Planck Institut f\"ur Microstruktur Physik, Weinberg 2, 06120 Halle, Germany}
\affil[3]{Max-Planck Institut f\"ur Chemie, Chemistry and Physics at High Pressures Group Postfach 3060, 55020 Mainz, Germany.}
\affil[4]{Institute of Theoretical and Computational Physics, Graz University of Technology, NAWI Graz, 8010 Graz, Austria}
\affil[5]{Dipartimento di Fisica Universit\`{a} degli Studi di L'Aquila and SPIN-CNR, I-67100 L'Aquila, Italy}
\affil[*]{jose.flores@unibas.ch}

\begin{abstract}
Pressure-induced superconductivity and structural phase transitions in phosphorous (P) 
are studied by resistivity measurements under pressures up to 170\,GPa and  
fully {\it ab initio} crystal structure and superconductivity calculations up to 350\,GPa. 
Two distinct superconducting transition temperature  (T$_{\textmd C }$) vs. pressure ($P$) trends at low pressure 
have been reported more than 30 years ago, and for the first time we are able to reproduce them and devise a consistent explanation founded on thermodynamically metastable phases of black-phosphorous. 
Our experimental and theoretical results form a single, consistent picture which not only provides a clear 
understanding of elemental P under pressure but also sheds light on the 
long-standing and unsolved {\it anomalous} superconductivity trend.  
Moreover, at higher pressures we predict a similar scenario of 
multiple metastable structures which coexist beyond their thermodynamical stability range.
Metastable phases of P experimentally accessible at pressures above 240\,GPa 
should exhibit \tc's as high as 15\,K, i.e. three times larger than the predicted value for the ground-state crystal structure.
We observe that all the metastable structures systematically exhibit larger transition temperatures than the ground-state ones,
indicating that the exploration of metastable phases represents a promising route to design materials with improved superconducting properties. 
\end{abstract}

\begin{document}
\flushbottom
\maketitle

\section*{Introduction}

The discovery that sulfur hydride (\sh) is superconductor with a record-breaking critical 
transition temperature (T$_{\textmd C }$) of 200\,K~\cite{DrozdovEremets_Nature2015}, has 
disproved a decades-long prejudice against high-\tc\ occurring conventionally~\cite{mazin2015superconductivity,Eliashberg}. 
This result demonstrated that extreme pressures represent a novel avenue to access new  
physical phenomena and exotic states of matter, which in the next years may lead to many surprises.  
Sulfur hydride is not an isolated example of conventional high-\tc\ superconductivity 
at high pressures, since a few months later also phosphine (\ph) has been observed 
to superconduct at transition temperatures as high as 100\,K at 200\,GPa~\cite{Drozdov_ph3_arxiv2015}. 

\begin{figure}[h!bt] 
\includegraphics[width=1.0\linewidth]{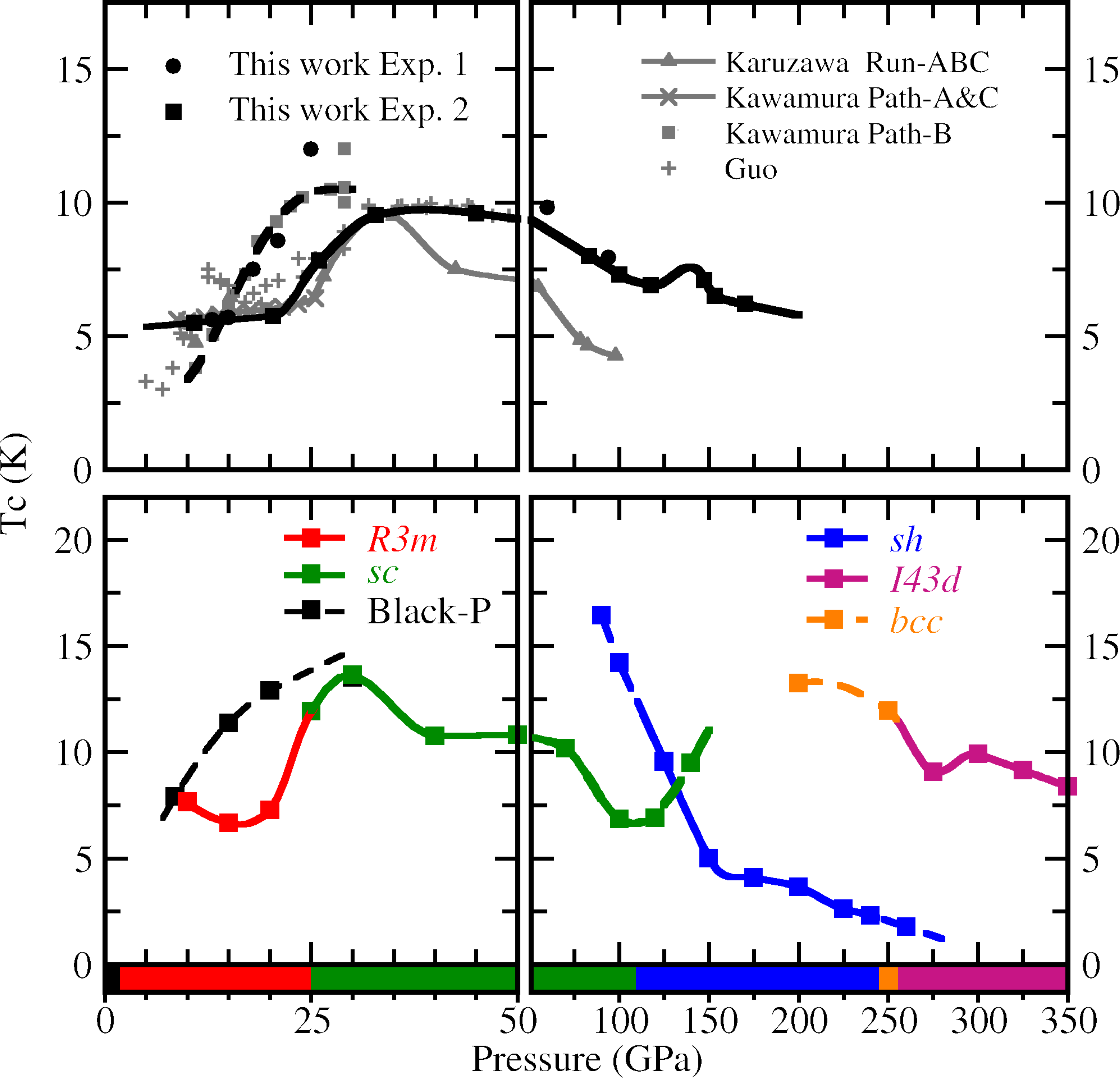} 
\caption{~Summary of all the critical temperatures observed experimentally~\cite{kawamura1985anomalous,karuzawa2002pressure,Guo_Ppressure2016} (top panels) 
and calculated theoretically in this work (bottom panels). 
In the experimental panels the dashed and full lines are a guidelines-to-the-eye and
used to distinguish the trend that we interpret as due to metastable black-P (dashed) 
from the energetically stable sequence of transformations (see text).  
The color-bar at the bottom indicates the sequence of calculated ground-state structures; 
the color-code is explained in the legend of the bottom panels.}
\label{fig:tc}
\end{figure}

The case of phosphine it is even more fascinating than that of \sh, 
whose crystal structure and superconducting phase diagram have been largely studied 
and unambiguously determined both experimentally~\cite{Ivan_Science-H2S-2016,Einaga_H3S-crystal_NatPhys-2016} 
and theoretically~\cite{Duan_SciRep2014,LiYanmingMa_JCP2014,SH_PRB-Mazin-2015,
Errea_anhaPRL2015,Flores-Livas_H3Se2016,ishikawa2016superconducting,Heil-Boeri_PRB2015,akashi_mangeli-phases}. 
In fact, the same theoretical methods which were used successfully to
 predict \sh, show that at high pressure \ph\ should decompose into its elemental components, phosphorus and hydrogen.
Since the {\em ab-initio} description of the thermodynamics of hydrides is in general
quite accurate, the most likely explanation is that the reported superconducting phase of \ph\ is 
to occur in a {\em metastable} structure of PH$_n$ ($n=1,2,3,4$) stabilized by a particular, 
nevertheless reproducible experimental condition~\cite{FloresLivas-PH3_PRB-2016,shamp_decomposition_2015}. 

Metastable phases, which can be accessed only under specific thermodynamical conditions,
play a major role in determining the high pressures properties 
in many compounds; in some cases, they can lead to complicated superconducting
phase diagrams, even for simple elements~\cite{sakata2011superconducting}.

The most prominent example is hydrogen, where the search for a possible metallic phase~\cite{Wigner_JCP1935}, 
which could be a room-temperature superconductor~\cite{Ashcroft_PRL1968} has been going on for decades.  
Actual hints for metalization have been recently reported in Ref.~\cite{eremets2016low} and were based 
on the temperature dependence of electrical resistance and disappearance of Raman spectra. 
However the reported metallization in Ref.~\cite{Dias_hydrogen_Science2017} observed for atomic 
hydrogen based on reflectance of the sample is still debated~\cite{goncharov2017comment,loubeyre2017comment,eremets2017comments}.  
What is certain is that the metallic phase involves metastable phases that are only accessible at high pressures 
and precise temperatures~\cite{brovman1972structure,eremets2011conductive,dalladay2016evidence,eremets2016low}. 

Although less spectacular in terms of critical temperature, phosphorus is another well-known
example of elemental compound, which becomes superconducting under pressure and has a rich and complicated phase diagram~\cite{buzea2004assembling}. 

In this work, we present a thorough investigation of the superconducting properties of phosphorus under pressure, 
including new experiments at high pressure and, for the first time, a fully \textit{ab-initio} characterization.     
We discovered that the structural metastability of phosphorus plays a crucial role in determining the superconducting 
phase diagram and unquestionably this is driven by pressure and temperature conditions.   
Thanks to our {\em ab-initio} study we were able to disentangle the nature of the superconducting properties of 
P metastable phases and more importantly to predict that these phases have larger transition temperatures than the putative ground state ones. 

\section*{Results}  

\subsection*{High-pressure Experiments} 
\begin{figure}[t!] \begin{center}
\includegraphics[width=1.0\linewidth]{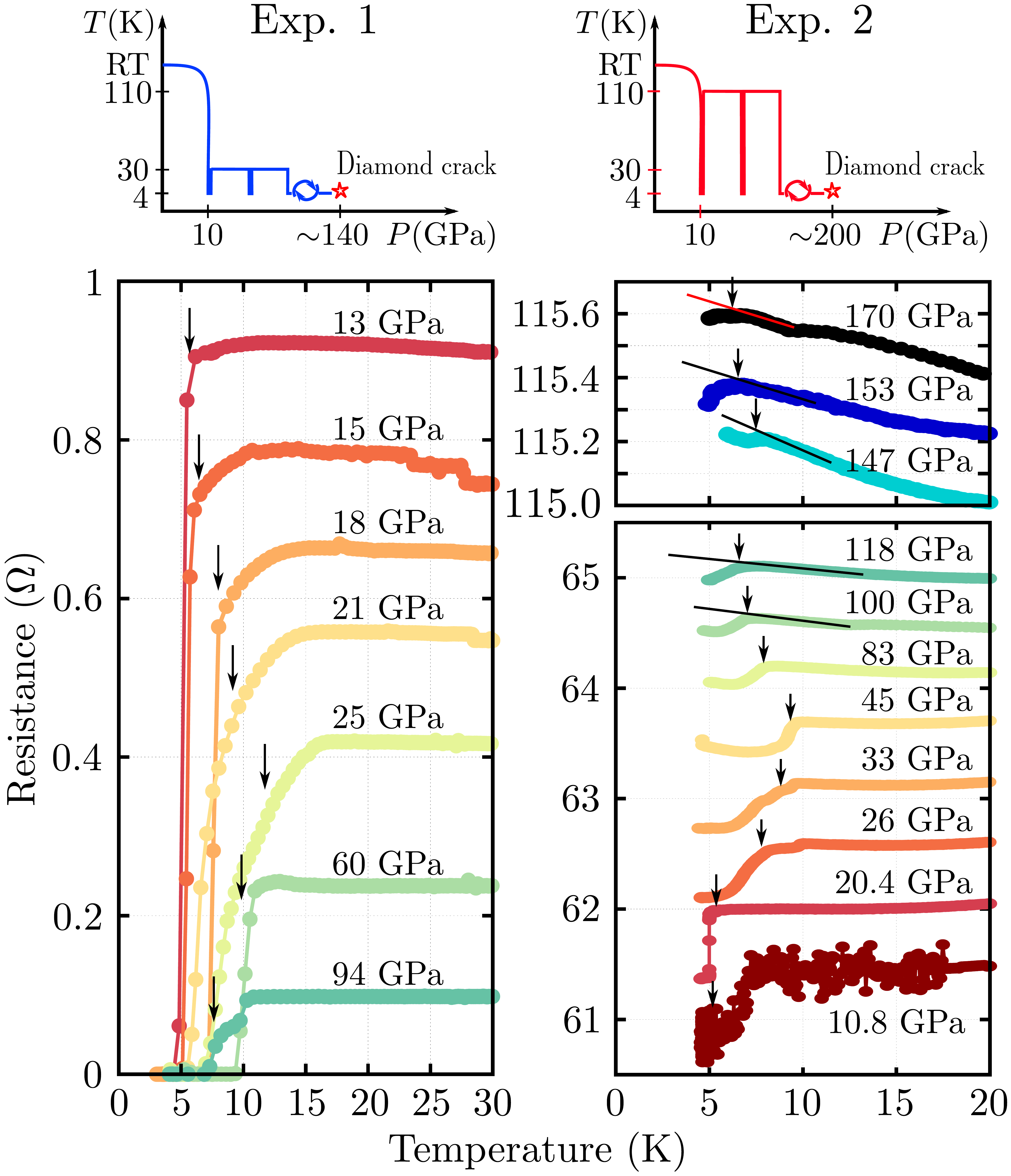} 
\caption{~Bottom panels show resistivity measurements as a function of temperature for different pressures conditions: 
Exp.~1 (top-left panel) the sample was cycled at low temperature ($<$~30\,K) and 
Exp.~2 (top-right panel) it was cycled up to high temperature (($<$~260\,K). 
The T$_{\textmd C }$ on-set for each pressure are marked with arrows.
The corresponding values are summarized in Fig.~\ref{fig:tc}.} 
\label{fig:resistance}
\end{center} \end{figure}

The superconducting phase diagram of P is intriguingly complicated; the
top panel of Fig.~\ref{fig:tc} compares the previous experimental results  
(gray symbols and lines) with our new experiments (black symbols and lines).  
The first accurate resistivity measurements on P under compression date back to 1985 
and were performed by Kawamura {\it et. al.}\cite{kawamura1985anomalous}.  
The authors reported for the first time an {\it anomalous} behavior in the superconducting transition 
temperatures up to pressures of 30\,GPa: the value of \tc\ was found to strongly depend 
on the $P-T$ path followed in experiment.
The problem was revisited by Karuzawa {\it et. al.}~\cite{karuzawa2002pressure} in early 2000's, 
with a second set of experiments in which transition temperatures were measured for 
pressures of up to 97\,GPa, producing a single \tc\ trend. 
More recently Guo {\it et. al.}~\cite{Guo_Ppressure2016} carried out Hall-electrical measurements  
under pressure up to 50\,GPa, and reported a non-monotonic trend of \tc\ at low pressures, 
and an anomaly due to a Lifshitz transition at $\sim 17$\,GPa.

Clearly, despite the effort and several experiments over decades, the explanation of the superconducting 
trends and more importantly the anomalous dependence on thermodynamic conditions remains unsolved.  
Given the polymorphism of elemental phosphorus already at ambient conditions, 
one could speculate that the anomalous \tc\ vs. $P$ trends are caused by the coexistence of metastable phases.  
Therefore, in the present work we design two different sets of experiments which specifically aim at accessing 
different phases (stable and metastable) following two $P-T$ thermodynamic paths, 
as schematically illustrated in top panels of Fig.\ref{fig:resistance}. 

In the first set, Exp.~1, (the measured \tc\ is reported as black circles in the top panel of 
Fig.~\ref{fig:tc}) the sample is constantly kept at low temperature while pressure is increased. 
In the second $P-T$ path (Exp.~2) the temperature is 
raised at higher temperatures when applying pressure and for resistivity measurements.  
In this figure bottom panels show the resistivity measurements at different pressures and \tc's are indicated with arrows. 
Clearly the \tc\ vs. $P$ behavior of the two datasets is rather different.
In the first set (black circles in Fig.~\ref{fig:tc}) a slow increase of \tc\ up to 25\,GPa is observed, 
while for the second set of experiments \tc\ sharply increases with pressure (black
squares in Fig.~\ref{fig:tc}). The two data sets merge at  25\,GPa, and a single trend 
is observed up to the highest common pressure measured (94\,GPa). 

\begin{figure}[htb]
\begin{center}
\includegraphics[width=0.95\linewidth]{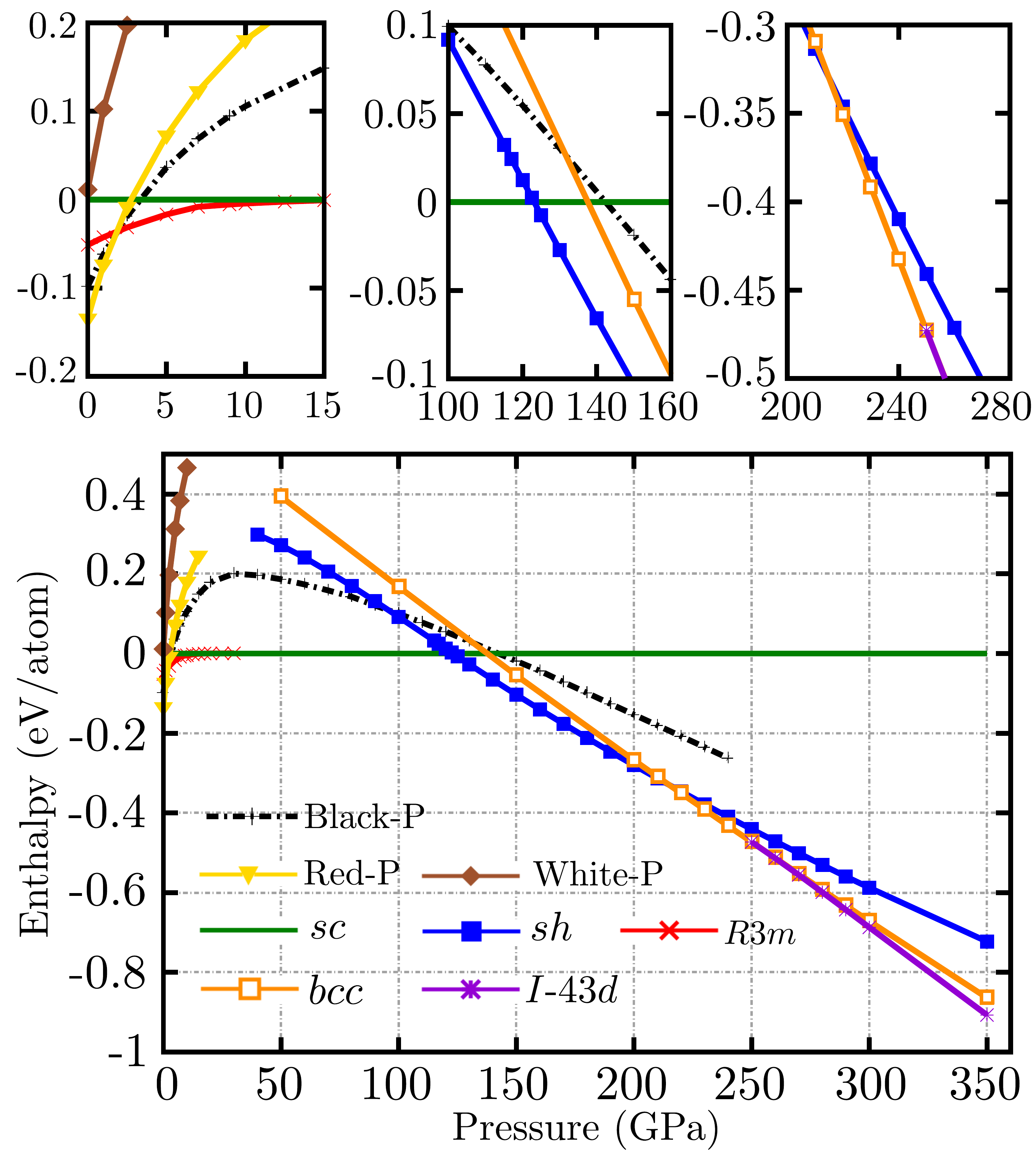} 
\caption{~Calculated enthalpy for different crystal structures of phosphorus 
w.r.t \simpcub\ as function of pressure. The top panels show an enlargement of 
three relevant pressure windows in which several structures are energetically 
competitive within orders of $\sim$~100\,meV (left), $\sim$~15\,meV (center) and $\sim$~5\,meV (right).} 
\label{fig:enthalpy}
\end{center}
\end{figure}

\subsection*{{\em Ab-initio} Phase diagram of P under pressure}

Fig.~\ref{fig:enthalpy} shows the computed enthalpy for different allotropes of phosphorus under pressure 
found using our crystal structure prediction method.  
The enthalpy difference is shown with respect the \simpcub\ ($Pm-3m$) phase. 
The lowest-enthalpy sequence of transitions, according to the calculations, is the following: 
Red-P (triclinic $P-1$)~\cite{roter_phosphor_2005} is stable at 0\,GPa and almost 
degenerate with black-P ($Cmca$)~\cite{hultgren1935atomic}, which is the experimentally observed phase; 
phase $P-II$ ($A7$-$R3m$) occurs from 3 to 16\,GPa. The simple cubic (\simpcub -$P-III$) 
lattice dominates for pressures up to 120\,GPa, where the simple hexagonal lattice 
(\hex -$P6mmm$) is stable up to $\sim$~225\,GPa. 
We find the \bcc\ ($Im-3m$) crystal stable from 225 to 250\,GPa and finally the last phase 
(\Iftd\ ) is stable from 250 to beyond 350\,GPa. 
The sequence of transitions is shown as a color bar at the bottom of Fig.~\ref{fig:tc}.

The three relevant pressure intervals, in which phase transitions take place, are highlighted 
in top panels of Fig.~\ref{fig:enthalpy}.  
The first one (left) shows the low-pressure regime. At ambient conditions phosphorus 
is known in at least three different allotropic forms: black-P, red-P and white-P. 
Experimentally, black-P is the most stable form~\cite{hultgren1935atomic} which 
transforms to the \rhomb\ phase~\cite{P-bcc_PRB}  
for pressure above 5\,GPa~\cite{jamieson1963crystal}. 
In our calculations, red-P is the ground state at zero pressure (although is almost degenerate with black-P), while
the stabilization energy of white-P is 130\,meV per atom higher (or equivalently $\sim$~1500\,K). 
The apparent disagreement is due to the fact that the standard Generalized Gradient Approximation (GGA) 
functional used in this work is inaccurate for layered (van-der-Waals bonded) or polymeric systems 
in predicting the exact structural sequence for different polymorphisms. 
However, although the Van-der-Waals interaction is rather important in the very low-pressure region 
of the phase diagram (0-5 GPa), it plays only a minor role at higher pressures, where superconductivity occurs. 
We find that black-P remains enthalpically competitive 
within a comparable order-of-magnitude ($~\sim$ 0.1 eV)  with other 
low-enthalpy phases up to pressures as large as 30\,GPa.
As will be discussed in detail in the next section, the meta-stability of this phase turns out 
to be fundamental to describe the experimental trend in \tc.

The second panel (top central) of Fig.~\ref{fig:enthalpy} shows the pressure interval 
in which the \hex$\to$\bcc\ transition occurs. 
In this window \simpcub, \hex\ ($P6mmm$), IM-$Cmmm$ (not shown), \bcc\ and, surprisingly, black-P (in a collapsed form) 
are  structures that are all accessible within a few tens of meV energy difference. This enthalpy landscape is consistent 
with  experimental evidences that in this pressure range the  \hex$\to$\bcc\ transition occurs via 
an intermediate incommensurate phase~\cite{Marques_P-PRB-2008}.  

A third pressure range, worth analyzing in more detail (top right panel), is where the \hex\ and \bcc\ 
enthalpy curves cross each other and the \bcc\ to \Iftd\ transformation occurs (\Iftd\ is a distorted form of \bcc). 
Note that in this case the enthalpy differences between all three phases are extremely small, i.e. within the computational accuracy, so that vibrational entropy corrections (not included) and stabilization of distorted complex structures~\cite{P-bcc_PRB,Sugimoto_Psuperlattice-PRB-2012} could in principle affect the energetic ranking 
of the structures and the transition pressures. 
However, these corrections will shift the transition pressure by no more than $\sim 5\,$ GPa, which is below 
the experimental errors to estimate the pressure ($\sim 10\,$ GPa). 

Indeed, our {\it ab initio} zero-temperature phase diagram is in substantial agreement with the 
sequence experimentally observed~\cite{jamieson1963crystal,P-bcc_PRB,Akahama_hexgonal_P_PRB1999,Christensen_BaIV-P_PRB2004,
Phosphorus_IV_PRL2006,Incommesurate_P-IV-PRL2007,Marques_P-PRB-2008,Toledano_a7structuresPRB-2008,Boulfelfel_P-lonepairs_PRB2012}, 
and is therefore a good starting point to calculate the superconducting critical temperatures as a function of the pressure.
Furthermore, our analysis has allowed us to characterize several of the structural transitions as first-order, 
i.e. with a discontinuous $P(V)$ behaviors (see Fig.~\ref{fig:a2f}); this 
 could lead to a possible path to stabilize metastable structure under suitable thermodynamic conditions.  

\subsection*{{\em Ab-initio} predicted superconducting temperatures} 

Using state-of-the-art Density Functional Theory for Superconductors (SCDFT) combined  with density-functional perturbation theory for the calculation of phonon dispersions and electron-phonon coupling
and linear response theory in the random phase approximation for the evaluation of electron-electron repulsion, 
we have computed the superconducting properties (anomalous density and critical temperature, \tc) for all the identified 
structures of P which are dynamically stable. The corresponding \tc's are
shown in the lower panel of Fig.~\ref{fig:tc}, as full (dashed) lines for ground-state (metastable) structures.

\begin{figure}[t!]
\begin{center}
\includegraphics[width=1.0\linewidth]{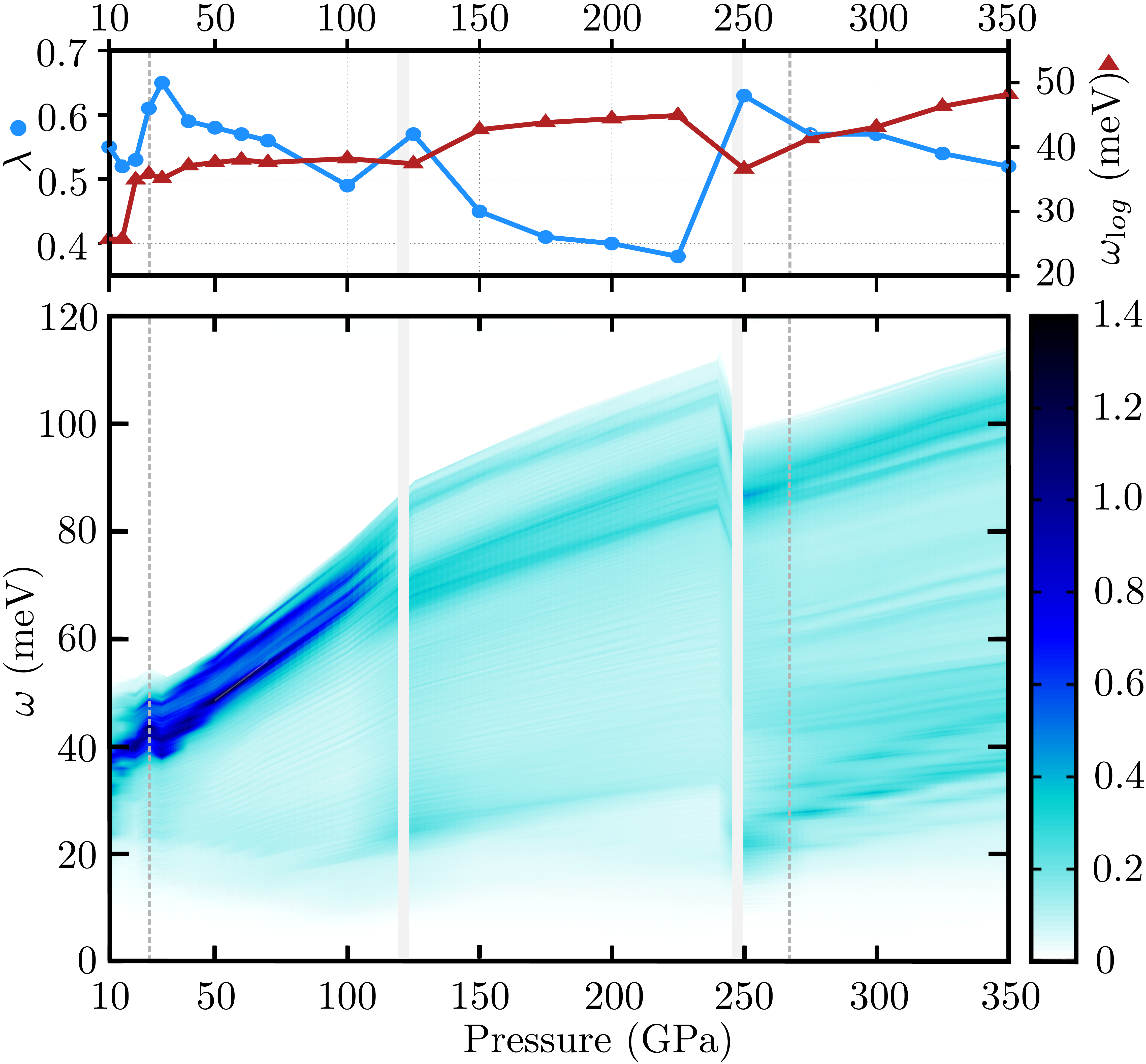} 
\caption{~The bottom panel shows the pressure-evolution of the Eliashberg function $\alpha F^2(\omega)$ 
(Eq.~\ref{eq:a2F}), for all the thermodynamically stable phases of phosphorus between 10 and 350\,GPa. 
Pressure are shown in abscissas, frequencies ($\omega$) in ordinates; the function value is given by the color-scale. 
White vertical gaps mark first-order phase transition and gray dashed lines mark second-order transitions. 
The top panel shows the electron-phonon coupling constant $\lambda$ (left $y$-axis) and the characteristic 
phonon energy $\omega_{\rm log}$ (right $y$-axis) obtained from the Eliashberg spectral functions.
(see Methods and Supplemental Materials).}
\label{fig:a2f}
\end{center}
\end{figure}

\paragraph{Low Pressures (0-50\,GPa)} 
As shown in the previous section, phosphorous has an extremely rich phase diagram of unique complexity; at ambient conditions 
of pressure and temperature its three most common polymorphs, black-P, red-P and white-P are semiconducting or insulating. 
While black-P and red-P achieve metalization within a few GPa (estimating the exact value would require calculating 
corrections to the band-gap beyond DFT),  white-P is still insulating at 7\,GPa with a band gap of about 1.4\,eV. 
Considering that DFT typically underestimates band gaps in insulators, this implies that white-P cannot be considered a candidate for the superconducting phase, and therefore will not be included in further investigations. 
Also red-P is excluded as a possible superconducting phase in the low-pressure range, because, although dynamically 
stable  for 10 and 15\,GPa, the calculated \tc\ is too low ($<$~1\,K) to account for any experimental evidence of superconductivity.
Upon further compression our calculations shown that red-P collapses to a simple cubic lattice.  
For black-P, on the contrary, the calculated \tc, depicted as black line in Fig.~\ref{fig:tc}, rapidly increases as a function of pressure.  This a consequence of the relatively strong electron-phonon coupling, which is consistent with previous 
predictions on doped black-P and phosphorene~\cite{SannaProfeta_BlackP_2Dmaterials_2016}. 
This superconducting structure of black-P, which remains energetically {\it metastable} 
and dynamically stable for pressures up to 20\,GPa,  with transition temperatures as high 
as $\simeq$~13\,K, is actually a modulated superstructure  of the true black-P, obtained by 
supercell relaxation along soft directions. Upon further compression the modulated black-P 
relaxes to the simple cubic lattice.  

However, at pressures in which superconductivity occurs (~$\sim$~10\,GPa) the ground-state structure is not black-P, 
but \rhomb-P, in good agreement with XRD measurements~\cite{Sugimoto_Psuperlattice-PRB-2012} and previous theoretical calculations~\cite{Cohen_LowP_P-PRB2013}. 
In this phase the calculated evolution of \tc\ is pressure-independent; \tc\  does remains constant 
$\sim$~7\,K (red-line in bottom panel in Fig.~\ref{fig:tc}). 
Noticeably, we find that the sudden increase in \tc\ for pressures 
above 20\,GPa occurs in correspondence to the \rhomb\ $\to$ \simpcub\ second-order phase transition. 
Our calculations show that this substantial increase in \tc\  is not originated, as one could expect,
from soft phonons inducing the structural transition which could consequently provide large electron-phonon coupling, 
but is instead triggered by an {\em electronic} Lifshitz transition~\cite{LifshitzTransition_JETP1960}. 
It does happen indeed that, upon increasing pressure, an additional band crosses the Fermi level, contributing to the 
electronic density of states (DOS) and providing a strong interband scattering channel that reinforces the Cooper pairing. 
An extensive analysis of this effect is provided in the Supplemental Material. 
The same behavior was recently predicted also in the superconducting properties of elemental sulfur 
under pressure~\cite{Monni_S_pressure_2017}, which remarkably has many common aspects with phosphorus. 

Once the Lifshitz transition is completed, \tc\ marginally decreases with pressure as the occupation of the 
additional band increases, enhancing the electronic screening of the electron-phonon coupling.
As a consequence \tc\ results rather featureless in the range between 40 to 100\,GPa, where the only relevant phase is
the \simpcub\ one, showing \tc's decreasing with pressure, in agreement with previous estimation~\cite{Cohen_LowP_P-PRB2013}. 
Considering only the thermodynamically stable structures, the predicted \tc\ follows closely the pressure dependence of the 
BCS-type coupling $\lambda$ (Eq.~\ref{eq:lambda}), reported in Fig.~\ref{fig:a2f} (blue curve in top panel). 
This is due to the combination of three factors: {\em first}, superconductivity is happening in the "weak-coupling limit", 
since the values of $\lambda$ are between 0.4 and 0.7. In this limit \tc\ is extremely sensitive to $\lambda$.  
{\em Second}, the characteristic phonon energy scale \omlog\ (Eq.~\ref{eq:omegalog}) and red line in the upper 
panel of Fig.~\ref{fig:a2f}) shows almost no pressure dependence,
as would occur in presence of phononic instabilities. 
{\em Third}, the repulsive Coulomb interaction between electrons in the Cooper pairs, represented by  the Coulomb pseudopotential 
parameter $\mu^*$~\cite{AllenMitrovic1983,MorelAnderson_1962},
is constant and $\simeq$ 0.1 for all structures and pressures, which is  rather typical for $sp$-bonded systems. 
More insight on the pressure-evolution of the characteristic $ep$ coupling parameters for the thermodynamically
stable phases can be obtained analyzing the pressure evolution of the electron-phonon spectral function $\alpha^2 F(\omega)$,
shown in the bottom panel of Fig.~\ref{fig:a2f}. In the plot, frequencies are shown in ordinates,
pressures in abscissas, and the color scale indicates the intensity of the $\alpha^2 F(\omega)$; 
strongly-coupled phonons appear as darker regions at the corresponding frequencies.
In general, the maximum phonon frequencies raise as pressure is increased. 
Up to around 20\,GPa, the coupling is concentrated in the high-frequency modes, while the distribution changes at higher pressures (25--50\,GPa) where it appears more evenly distributed.  
This means that, despite the pronounced hardening of the maximum phonon frequencies with pressure, 
the center of mass of the spectrum remains almost constant, 
and there are no signatures of lattice instabilities induced by electron-phonon coupling. 

In contrast to the weak-coupling behavior of thermodynamically stable 
phases, {\em metastable} phases (black-P, \simpcub, \hex\ and $bcc$) 
exhibit a marked strong-coupling behavior at the limit of their dynamical stability range. 
For these systems, upon approaching their structural transition, a considerable part of the phonon spectral weight 
is shifted to low frequencies, consequently increasing $\lambda$ and a simultaneously decreasing \omlog. 
Further details, including the calculated $\alpha^2 F(\omega)$, $\lambda$, \omlog\ and the Coulomb screening parameters 
for all phases discussed in this work can be found in the Supplemental Material.

\paragraph{High Pressures (50-350\,GPa)} 
We will now analyze the high-pressure range, for which our calculated \tc's are 
summarized in the right bottom panel of Fig.~\ref{fig:tc}. 
In the \simpcub\ phase, \tc\ continues to decrease until the \simpcub\ to \hex\ transition occurs.  
Since this is a transition of first order, as the one from black-P to  \simpcub, 
we investigated the metastability of the two phases across their thermodynamic boundaries. 
For higher and lower pressures and for both systems our calculations predicts dynamically stable structures: 
\simpcub\ up to 150\,GPa and \hex\ down to 80\,GPa. 
Away from their thermodynamical phase boundaries, \tc\ varies rapidly in both systems and this is due to
a phonon softening that lowers the phonon energy and increases the coupling.  
From 150 to 250 GPa, the \tc\ of \hex-P decreases steadily below 3\,K,
and this phase is therefore moderately interesting for superconductivity. 
On the other hand,
the \bcc/\Iftd\ sequence of structures, stable above 250 GPa, is more promising. 
In fact, these phases show a larger \tc, well above 10\,K. 
It is also worth mentioning that, between 220 and 250\,GPa the \bcc\
structure which is thermodynamically metastable, exhibits transition temperatures 
three times larger than those calculated for the \hex\ structure. 

\section*{Discussion} 

One of the crucial points of this work is to identify metastability as a distinctive 
feature of the phase diagram of phosphorus. In this element, the presence of several discontinuous phase transitions, 
not directly triggered by phononic instabilities, implies the coexistence of several metastable 
phases with distinct superconducting properties across the phase diagram. 
This means that the initial conditions and the $P-T$ path followed in the experiments largely determine 
the resulting crystal structure, coexistence of phases and the possibility of observing high-\tc\ metastable phases. 

Using high-pressure experiments and first-principles calculations we are able to
draw a consistent picture of the full phase diagram up to 350\,GPa. 
We propose that the coexistence of two distinct superconducting phases of 
phosphorus in the 10--30\,GPa pressure window successfully accounts for the the "anomalous" \tc vs. $P$ trend 
first observed by Kawamura and coworkers~\cite{kawamura1985anomalous}. 
Furthermore, we identified theoretically two other pressure regions 
in which superconducting 
\tc\ anomalies could be observed depending on the experimental conditions.

We first discuss the 10--30\,GPa pressure interval, for which several experimental data are available.
Here, we hypothesize that three phases actually play a role in the superconducting phase diagram, 
namely black-P, $R3m$ and \simpcub, and that the bifurcation in \tc\ seen by experiments is due to the 
coexistence the ground-state low-\tc\ \rhomb-P and high-\tc\ metastable black-P
phases, which both collapse to \simpcub at $P \sim 20 GPa$. 
Our hypothesis  perfectly accounts for the existing literature results of Kawamura $et al.$,~\cite{kawamura1985anomalous},
in which the pressure cell was initially loaded with black-P,  and 
pressure was subsequently increased either keeping the system at {\em low temperature} (Kawamura Path B in Fig.~\ref{fig:tc}) 
-- or following a {\it room temperature} path (Path A and C). The first experimental procedure leads to the highest \tc, 
and is remarkably well reproduced by calculations for black-P, while the second leads to a low \tc, and
matches the calculated trend for \rhomb\ phosphorus. 

The two new sets of experiments in this work were explicitly 
designed to reproduce the distinct \tc\ trends, and to test our interpretation in terms of \rhomb\ and black phosphorus. 
In contrast to previous experiments, in this work the cell was loaded with red-P at room temperature. 
In the first set (Exp.~1), the samples were kept at {\em low temperatures} during the whole pressure run.
The first pressure measured was high enough ($\sim$~5\,GPa) to ensure a complete transition 
of red-P to black-P; according to our calculations, the transition should occur at $\sim $ 2.5\,GPa, 
and black-P should survive as a metastable phase for pressures  well above 20\,GPa, where it collapses to the \simpcub. 
On the contrary, following the second path in the $P-T$ phase space (Exp.~2),  
where annealing cycles allow to reach the thermodynamically ground-state structures, 
will stabilize the \rhomb\ phase between 5 and $\sim ~$20 GPa. Upon further compression the \rhomb\ 
structure will transform continuously to \simpcub\ as our crystal prediction method confirmed.  
The measured \tc\ vs. $P$ trends for the two paths closely reproduce
the corresponding measurements of Kawamura {\em et al.},
and are in excellent agreement with our theoretical predictions,
 supporting our interpretation that the high-\tc\ phase is metastable black-P.
An alternative  explanation would be to ascribe the high \tc\ curve in Exp.~1 to a metastable red-P phase. 
We can safely rule out this hypothesis because, according to 
our calculations, after the metalization red-P has a negligible \tc\ ($< ~$1\,K). 

It is worth to comment the recent measurements  by Guo \textit{et al.}~\cite{Guo_Ppressure2016}. 
They extend up to 50\,GPa, and are generally consistent with previous measurements and ours, but seem to contradict our understanding of the low-pressure range. 
In fact, the authors report  a gradual increase in \tc\ from 5 to 10\,GPa followed by an abrupt jump, 
which does not have a straightforward interpretation.
Our theoretical results suggest that a possible explanation for this phenomenon is the existence of a 
mixed black-P--$R3m$ (O+R) phase (suggested by the same authors) 
which prevents a clear observation of the two distinct \tc\ trends, 
and is probably responsible for the oscillations in \tc\ observed up 
to $\sim$25\,GPa, where the samples  enter to the \simpcub\ phase and Guo's data 
merge with our measurements and older data.

Besides the low-pressure region already explored by Kawamura $et al.$, we predict two other ranges of 
pressure in which different crystal structures 
are energetically accessible, and experimentally measurable:
the first is around $\sim$~110\,GPa, and a second is above $\sim$~220\,GPa.  
In the first case, unfortunately the two competing phases have very similar critical temperatures 
(see \tc\ for $hex$ and $sc$ $\sim$ 120\,GPa), and are thus hard to distinguish experimentally.
Moreover, phonons are softening close to the transition,
coming both from low and high pressures, indicating that it 
could be difficult to stabilize metastable structures and that
complex disordered or modulated structures may form~\cite{marques_origin_2008}. 
In the second case, the \tc's of the ground-state and metastable structures (which according to our predictions 
is of comparable energetics at the relevant pressures) differ by a factor of three 
(see Fig.~\ref{fig:enthalpy} and \tc's for $sh$, $bcc$ and \Iftd\ above $\sim$ 225\,GPa). 
Therefore, these two distinct \tc\ vs. $P$ trends should be easily discernible by experiments. 
However, no superconductivity measurements have been reported yet at these pressures, which was 
indeed unaccessible also for our  current experimental setup. 
Nevertheless we note that, remarkably, our theoretical prediction of phase coexistence is
consistent with recent XRD experiments~\cite{Sugimoto_Psuperlattice-PRB-2012}. 
A possible path to achieve the synthesis of these metastable phases would be to start from the ground state structure 
at high pressures \Iftd\ ($\lesssim$270\,GPa), cycle the sample trough temperatures high enough to overcome the
energetic barrier to the \hex\ phase, and then cool  
the sample {\it slowly} releasing the pressure  down to 230\,GPa. 

In summary, we have conducted a systematic theoretical and experimental investigation of elemental phosphorous under pressure. 
We have shown an excellent agreement between our experimental and theoretical results, which allow us 
not only to reconcile previous unexplained observed anomalies, but also to shed light onto the complex behavior of this element, which has the tendency to form many polymorphs and that differ substantially in their electronic and superconducting properties and can coexist in metastable forms in different pressure ranges. 
A similar behavior has been reported in a wide variety of conventional superconductors~\cite{BACI}, including simple elements~\cite{Ca_competing_phases}.  
The selective stabilization of metastable phases may represent, in the future, a viable strategy to improve the superconducting properties 
of conventional superconductors. For this, it is essential to assess the relative accuracy of 
experiments and calculations in this respect.

\section*{Methods} \label{sec:Methods}

\paragraph{Crystal phase diagram exploration}
To sample the enthalpy landscape we employed the minima hopping method (MHM)~\cite{Goedecker_2004} with 
unit cells  of up to 8 atoms for selected pressures in the range of 0 to 350\,GPa.  
This method has been successfully used for global geometry optimization in a large variety of applications~\cite{Amsler_2010,MA_JAFL,Disilane_JAFL,LiAlH_Maxmotif,BJAFL_PRB2012}, including 
superconducting materials at high pressure~\cite{Disilane_JAFL}. The MHM was designed to thoroughly scan the low-lying enthalpy landscape of any compound and identify 
stable phases by performing consecutive short molecular dynamics escape steps followed by local geometry relaxations. 
The enthalpy surface is mapped out efficiently by aligning the initial molecular dynamics velocities approximately along 
soft-mode directions~\cite{roy_2009,sicher_efficient_2011}, thus exploiting the Bell-Evans-Polanyi~\cite{jensen_introduction_2011} 
principle to steer the search towards low energy structures. 
Energy, atomic forces and stresses were evaluated at the density functional theory (DFT) 
level with the Perdew-Burke-Erzernhof (PBE)~\cite{PBE96} parametrization to the exchange-correlation functional. 
A plane wave basis-set with a high cutoff energy of 1000\,eV was used to expand the wave-function together with 
the projector augmented wave (PAW) method as implemented in the Vienna Ab Initio Simulation Package~{\sc vasp}~\cite{VASP_Kresse}. 
Geometry relaxations were performed with tight convergence criteria such that the forces on the atoms were less than 2~meV/\AA~and 
the stresses were less than 0.1~eV/\AA$^3$.  
We have reproduced all the experimentally-known phases of P and other low-lying phases, 
except for white-P and red-P, for which larger supercell calculations are necessary to describe the structure. 

\paragraph{Coupling and superconductivity calculations}
All superconductivity calculations are performed within Density Functional Theory for Superconductors (SCDFT)\cite{OGK_SCDFT_PRL1988,Lueders_SCDFT_PRB2005}. 
The approximation used have been described in previous works~\cite{Flores-Sanna_PRBhoneycombs,Flores-Livas_H3Se2016,FloresLivas-PH3_PRB-2016,Linscheid_localOP_PRL2015,Monni_S_pressure_2017}.
The pairing mechanism is due to the combined effect of electron-phonon coupling within DFT Kohn-Sham theory~\cite{DFPT_S.Baroni} as implemented in the {\sc Quantum Espresso} code, and electronic screening computed in the static random-phase-approximation~\cite{Massidda_SUST_CoulombSCDFT_2009}.
This allows us to calculate \tc\ completely {\em ab-initio}, without introducing any
empirical parameter, such as the  $\mu^*$ Coulomb pseudopotential usually used so solve the Eliashberg equations. Still, we can estimate an effective $\mu^*$ (reported in the Supplemental Material), fitting the fully {\em ab-initio} \tc\ with the  Allen-Dynes-McMillan formula~\cite{AllenDynes_PRB1975}.
The electron-phonon coupling at the Fermi energy is described in the isotropic approximation by the Eliashberg spectral functions~\cite{AllenMitrovic1983}, defined as:
\begin{equation}
 \alpha^2 F(\omega) = \frac{1}{N_{E_F}} \sum \limits_{\mathbf{k} \mathbf{q},\nu} |g_{\mathbf{k},\mathbf{k}+\mathbf{q},\nu}|^2 \delta(\epsilon_\mathbf{k}) \delta(\epsilon_{\mathbf{k}+\mathbf{q}}) \delta(\omega-\omega_{\mathbf{q},\nu})~, \label{eq:a2F}
\end{equation}
where $N_{E_F}$ is the DOS at the Fermi level, $\omega_{\mathbf{q},\nu}$ is the phonon frequency of mode $\nu$ at wavevector $\mathbf{q}$ and 
$|g_{\mathbf{k},\mathbf{k}+\mathbf{q},\nu}|$ is the electron-phonon matrix element between two electronic states with momenta $\mathbf{k}$ and $\mathbf{k+q}$.
All computed $\alpha^2 F(\omega)$ are collected in Fig.~\ref{fig:a2f}. 
Anisotropy effects have been estimated to be irrelevant in the calculation of \tc\ and are neglected in this work. Two significant moments of the Eliashberg function $\lambda$ and \omlog, defined as:
\begin{eqnarray}
   \lambda&=&2\int\frac{\alpha^2 F\left(\omega\right)}{\omega}d\omega \label{eq:lambda}\\
   \omega_{\textmd log}&=&\exp\left[\frac{2}{\lambda}\int\alpha^2 F\left(\omega\right)\frac{\ln\left(\omega\right)}{\omega}d\omega \right] \label{eq:omegalog}
\end{eqnarray} 
express, respectively, the electron-phonon coupling and the effective phononic energy.

Core atomic states are described in the norm-conserving pseudo-potential approximation; 
valence state are described by a plane-wave basis set with an energy cut-off 80~Ry. 
Since convergence checks have been performed independently on each phase, Brillouin zone integration 
is done with different sets of $k$-points in each crystalline structure, ranging from a minimum 
of 500 $k$-point per unit reciprocal volume (Bohr$^3$) up to about 3,000 $k$-points per unit volume 
for electronic integration and about one fourth of this density for phononic sampling. 
The strict convergence criteria ensure that the numerical error in the solution of the 
SCDFT equations is small as compared to the intrinsic error-bar on the available functionals.

\paragraph{Experimental procedure}
High pressure electrical measurements were carried out using a diamond anvil cell (DAC) 
with an anvil tip diameter of 200--300\,mm  bevelled at 7--8 degrees and with culet surface between 40--80\,mm. 
Four Ti electrodes were sputtered on the diamond anvil for the first experiment (Exp.~1) and 
three Ti electrodes for the second experiment (Exp.~2). The electrodes were capped with Au to prevent oxidation of the Ti 
(a zero contribution by the diamond surface to the conductivity was checked).   
An insulating gasket of Teflon was used to separate the metallic gasket from the electrodes. 
Red phosphorous was loaded in the DAC at ambient temperature and clamped at these conditions. 
The first pressure point after clamped was already about 3--5\,GPa in both experiments. 
Then, the DAC was placed into a cryostat and cooled down to measure the first point of 
\tc\ at 13\,GPa for the first experiment (Exp.~1) and 11\,GPa for the second experiment (Exp.~2). 
The pressure was determined by a diamond edge scale at low temperatures using Raman spectra and we monitored 
any possible Raman signal from the samples. 
We used the 632.8\,nm line of a He--Ne laser to excite the Raman spectra measured with a 
Raman spectrometer equipped with a nitrogen-cooled CCD and notch filters with resolution better than 2\,cm$^{-1}$.
Two different conditions were tested in our experiments, Exp.~1 and Exp.~2 as summarized in Fig.~\ref{fig:resistance}, 
with the only difference being the increased temperature at the beginning of the experiment. 
All the resistance measurements were done in increasing pressure; we could not perform measurements under 
decompression due to the diamond culet crack. 

\bibliography{main}

\section*{Acknowledgments}
The authors acknowledge the hospitality of the {\em cini-Sardegna} meeting, where part of this work was written. 
J.A.F.-L. acknowledges computational resources under the project 
(s499 and s707) from the Swiss National Supercomputing Center (CSCS) in Lugano. 
A.D. and M.E acknowledge the European Research Council with the 2010-Advanced Grant 267777. 

\section*{Author contributions statement}
J.A.F-L., A.S., L.B. and G.P. conceived the main idea of the research project. 
A.D. and M.E. performed the experiments. 
J.A.F-L. and  A.S. executed the ab-initio calculations.  
J.A.F-L., A.S., A.D., L.B. and G.P. analyzed the results and wrote the manuscript. 
M.E. and S.G. reviewed the manuscript. 

\section*{Additional information}
The authors declare no competing financial interests. 

\end{document}